\documentclass[aps,prl,floats,twocolumn,showpacs]{revtex4}

\usepackage{graphicx}
\usepackage{rotate}
\usepackage{epsfig}
\usepackage{latexsym,amsmath}

\begin{document}

\title{On the universality of the scaling of fluctuations in traffic on complex networks}

\author{Jordi Duch}

\affiliation{Departament d'Enginyeria Inform{\`a}tica i
Matem{\`a}tiques,
  Universitat Rovira i Virgili, 43007 Tarragona, Spain}

\author{Alex Arenas}

\affiliation{Departament d'Enginyeria Inform{\`a}tica i
Matem{\`a}tiques,
  Universitat Rovira i Virgili, 43007 Tarragona, Spain}

\begin{abstract}
We study the scaling of fluctuations with the mean of traffic in complex networks using a model where the arrival and departure of ``packets" follow exponential distributions, and the processing capability of nodes is either unlimited or finite. The model presents a wide variety of exponents between $1/2$ and $1$ for this scaling, revealing their dependence on the few parameters considered, and questioning the existence of universality classes. We also report the experimental scaling of the fluctuations in the Internet for the Abilene backbone network. We found scaling exponents between $0.71$ and $0.86$  that do not fit with the exponent $1/2$ reported in the literature.

\end{abstract}

\maketitle

Recently, the theory of complex networks has started to cope with the problem of dynamics on networks. After much work devoted to the understanding of the network topology \cite{revnewman}, the physics community has begun to develop models to uncover the phenomena observed in the dynamics on complex networks. Some stylized models of traffic flow in complex networks \cite{prlnostre, tadic, yamir, zhao, singh, goh,santos,munoz, ashton, barthelemy} can be used to gain intuition about complex networks dynamics, and
to determine the leading parameters of the dynamic processes related to the network topology. 

The main results obtained up to now concerning traffic flow in
complex networks, are related to the  determination of  bounds for
this flow to become congested. Nevertheless, traffic on real
complex networks, as for example the Internet, is not driven
by congestion processes but by large fluctuations of the "normal"
traffic behavior. In the case of the Internet, the understanding of
the physical laws governing the nature of 
traffic is crucial because its implications in design, control and
speed of the whole network \cite{leland93}. 

In a couple of recent
articles, Menezes and Barabasi propose a model to understand the origin of
fluctuations in traffic processes in a number of real world systems, including the
Internet, the world wide web, and highway networks
\cite{menezes04a,menezes04b}. All of
these systems can be represented at an abstract level as networks in
which packets travel from one node to another, packets being real data packets or bits in the Internet, files in the world wide web, and vehicles in road networks. 
In particular, the authors considered the relationship between the average number of packets
$\langle f_i \rangle$ processed by nodes during a certain time
interval, and the standard deviation $\sigma_i$ of this quantity.
They find that there are two classes of universality in this
relationship for real systems. In the Internet, $\sigma$ scales as $\langle f
\rangle^{1/2}$, whereas $\sigma$ scales as $\langle f \rangle$ for the world
wide web and highway networks. Based on a stylized model of random walkers throughout the network, they conclude that this difference is due to the
fact that the dynamics of the Internet is dominated by ``internal
noise'' whereas the dynamics of the world wide web and highway
networks is dominated by the demands of users, that is ``external
noise''.  In the abstraction process proposed by the authors, they
overlook what is probably one of the most important factors in the dynamics
of traffic on networks---the limited capacity of nodes to handle
packets simultaneously, which results in packet-packet interactions
and, eventually, in large fluctuations or even network congestion
\cite{prlnostre, tadic}.

In this Letter we show that simple considerations regarding the persistence of packets flowing the network, the limitation of nodes to handle information, and the time window where statistics are performed, account for different scalings of the fluctuations in traffic on complex networks. The main results obtained are: (i) Maintaining the total traffic on the network constant, different scaling laws arise depending on the relation between the ratio of packet input and the steps these packets perform before disappear; (ii) The time window affects the scaling exponent of the fluctuations in such a way that, for a small enough time window the scaling trivially satisfies $\sigma\sim \langle f \rangle^{1/2}$ always, no matter the dynamic process. When the time window is large enough, the rest of parameters will provide the precise scaling between $\alpha=1/2$ and $\alpha=1$ where $\alpha$ refers to the scaling exponent $\sigma\sim \langle f \rangle^\alpha$; (iii) The effect of the packet-packet interaction (queue system) account for different scaling exponents as well; (iv) We find that within this framework there is not enough evidence for deriving universality classes. We have checked the scaling for data flowing on the Abilene backbone network, and show that the scaling exponent is different from $1/2$.

To understand the origin of the scaling relations for the
fluctuations in networks, let us consider the behavior
of a single node---for example, a toll plaza in a highway---trying
to satisfy demands from users---vehicles arriving to the toll--. As we
learn from queueing theory \cite{allen}, two stochastic processes
fully determine the behavior of the node: (i) the arrival process by
which new packets arrive to the node, and (ii) the service process
by which the node satisfies the demands of the users, that is,
forwards the packets. The most common queue model corresponds to the
M/M/1 queueing system, where the randomness of the packets
generation assumes a random (Poisson) arrival pattern \cite{footnote1} and the
service distribution assumes a random (exponential) time \cite{footnote2}.

Taking into account these considerations, we propose to model the
traffic process in a complex network of $N$ nodes as $N$ queue systems of type M/M/1, and a random walk simulation of packets movement on the network. The arrival process of packets to the
network is controlled by a Poisson distribution with parameter
$\lambda$, each packet enters the network at a random selected node.
Once the packet arrives to the node enters a queue. The delivery of
the packets in the queue is controlled by an exponential
distribution of service times with parameter $\mu$. In our model, the packets will
perform $S$ random steps in the network before disappearing, being then $S$ a measure of the persistence of packets in the network. This dynamics is performed in continuous time, assuming that the time expended by packets traveling through a link is negligible. 

The system achieves a stationary state whenever the arrival rate of packets at each
node is lower than or equal to the delivery rate, otherwise the system congests. The arrival rate at each node $i$ is topology dependent
and follows a distribution whose mean is $\lambda^{ef}_{i}={\cal B}_{i}
\lambda$ where ${\cal B}_{i}$ is the algorithmic betweenness of node $i$. ${\cal B}_{i}$ is
defined as the relative number of paths in the network that go through node $i$ given a specific routing
algorithm \cite{prlnostre}. As a direct consequence, the node with
maximum algorithmic betweenness ${\cal B}_*$ determines the onset of
congestion. We will focus on the average number of packets
$\langle f_i \rangle$ processed by nodes during a certain time
interval $P$, and the standard deviation $\sigma_i$ of this
quantity. 

\begin{figure}[t]
  \epsfig{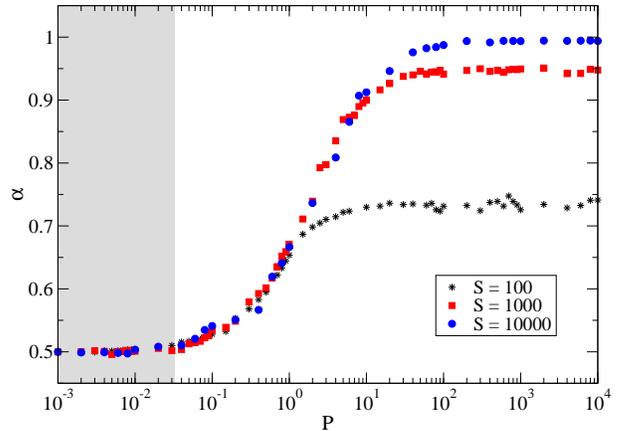}
  \caption{(Color online) Value of the exponent $\alpha$ versus the time window of size $P$, for a fixed
$\lambda^{ef}_*=1/3$ and different values of the persistence of packets in the network $S$. The shadowed area highlights the region of $P$ in which the exponent $\alpha=1/2$ always appears. The simulation is performed in a scale-free network with exponent for the degree distribution $\gamma = 3$ of 1000 nodes \cite{albert-barabasi}. We have observed the same results for larger SF networks at a subset of values of $P$, however the computational cost for the whole set of $P$ values used in the plot becomes prohibitive.}
  \label{periode}
\end{figure}

Selecting a value of $P \ll 1/\lambda^{ef}_{*}=1/({\cal B}_{*}\lambda)$,
we will always observe the scaling $\sigma\sim \langle f
\rangle^{1/2}$, regardless of other parameters. Due to the value of $P$ selected, the nodes will deliver either one packet or none, at each time interval. Suppose that during a
number $n_1$ of intervals of length $P$ the node deliver a packet
whereas it does not deliver during a number of intervals $n_0
=n-n_1$, where n is the number of samples for the statistics. In
this situation we also have $n_0\gg n_1$. Therefore, the average and
the standard deviation read

\begin{align}
    \langle f \rangle = n_1/n \\\nonumber
    \sigma = [\frac{1}{n}\left[n_1(1-\langle f \rangle)^2+n_0\langle f \rangle^2\right]]^{1/2}
\end{align}

\noindent which can be simplified to

\begin{equation}
    \sigma = [(1-\langle f \rangle)\langle f \rangle]^{1/2}
\end{equation}

But, in the current scenario, the average flow is $\langle f \rangle
\ll 1$ and then we recover the $\sigma\sim \langle f \rangle^{1/2}$
scaling law. Otherwise, this argument cannot be applied, and the
scaling value will be influenced by the rest of parameters of the
model.

In Fig.\ref{periode} we show the behavior of the scaling exponent
$\alpha$ as a function of the time intervals length $P$ over which the averages were taken, for a fixed
$\lambda^{ef}_*=1/3$. We observe (shadowed area) that the exponent is
always $1/2$ when the interval length is small enough. Indeed, from
the data used the exponent $1/2$ stands for values of
$P\lesssim0.01/\lambda^{ef}_*$.

Let us now assume that the sampling of the data is performed at
intervals of length $P\gg 1/\lambda^{ef}_{*}$. In this case, we
expect the scaling of fluctuations in the system,
beyond the effect of the sampling process, to be revealed. We analyze the behavior
of the system varying the rate of
injection of packets into the system $\lambda$ and the number of
steps $S$ each node performs before it disappears. We first
consider that the service rate $\mu\rightarrow
\infty$. In this case, the effect of queues is minimized and then no
interaction between packets is accounted for. The total traffic ${\cal T}$, number of
packets flowing through the network per unit time, is determined by
the Poisson process with mean $\langle {\cal T} \rangle= \lambda S $.

\begin{figure}[t]
 \epsfig{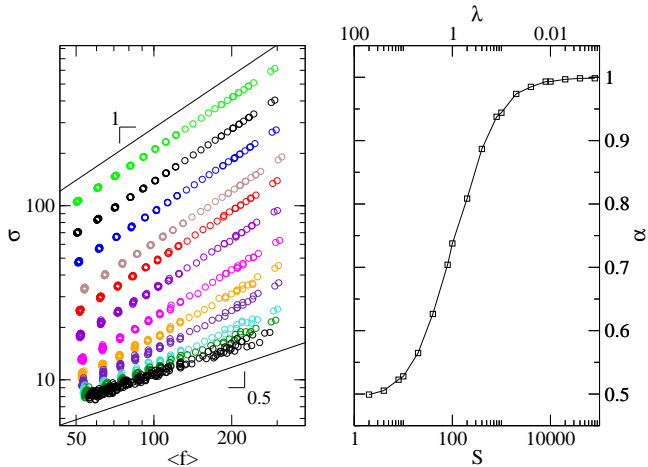}
  \caption{(Color online) Left: Plot $\sigma$ versus $\langle f \rangle$ for
  different realization of $\lambda$ and $S$ maintaining its product constant. The simulation is performed in a scale-free network with exponent for the degree distribution $\gamma = 3$ of 1000 nodes \cite{albert-barabasi}. We fixed ${\cal T}= \lambda S = 100$. Right: Plot of the $\alpha$ exponent for $\lambda S=100$. Other values of  $\lambda S$ have produced equivalent results, shifted to a different region of $\langle f \rangle$.}
\label{pasos}
\end{figure}

Keeping the total traffic mean $\langle \cal T \rangle$ fixed, we can control the
variability of the local traffic incoming to a node by varying the values
of $\lambda$ and $S$ proportionally. In Fig \ref{pasos}. we show the scaling
exponent transition between $\alpha=1/2$ and $\alpha=1$. This plot
recovers the results depicted in \cite{menezes04a}, although the
explanation should be reconsidered in the new scenario. The
transition of exponent from $\alpha=1/2$ to $\alpha=1$ is obtained here
simply by increasing the number of steps $S$ the packet performs on the
network while maintaining the mean value of the total traffic (i.e.
decreasing proportionally the injection ratio $\lambda$). This contradict the interpretation in \cite{menezes04a} because increasing the
number of steps in the network increases "internal
traffic" while decreasing the injection of packets is a decrement of
"external" traffic in this scenario. Nevertheless both results are
coherent at this point concerning the scaling of fluctuations. Our interpretation of this transition is the following: for the same total traffic on the network, the nature of fluctuations is related to the number of steps $S$ the packets perform on the network. When the number of steps is small enough the behavior of fluctuations is akin a random deposition process independent of the topology of the network, $\lambda^{ef}_{i} \approx \lambda$. When the number of steps in the network grows, the topology induces dynamical correlations that affect the scaling of fluctuations via the algorithmic betweenness, $\lambda^{ef}_{i} \approx \lambda {\cal B}_i$.    

\begin{figure}[t]
  \epsfig{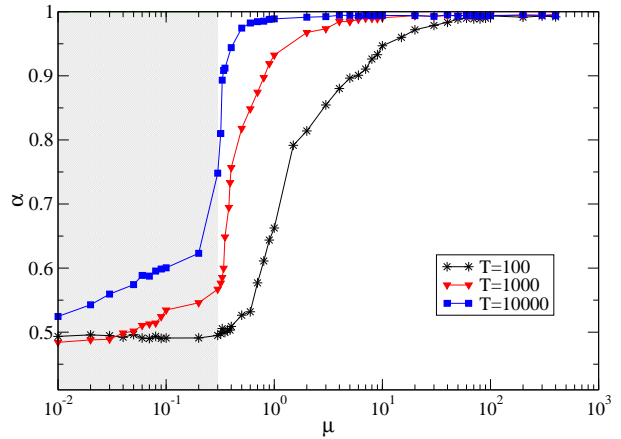}
  \caption{(Color online) Scaling exponent $\alpha$ as a function of the time service $\mu$, for three different time window sizes, and for $\lambda^{ef}_{*}=1/3$. Shadowed area highlights the region where congestion starts at nodes with $\lambda^{ef}_{*}=1/3$ .}
\label{mu}
\end{figure}

We extend the simple model where queues are neglected, to the
more realistic situation when queues are persistent. The
introduction of queues in the system, in our model, is controlled by
the parameter $\mu$ (rate of service). The possible values of $\mu$
are constrained by the onset of congestion i.e. $\mu
> \lambda^{ef}_{*}$, otherwise congestion appears at those nodes with ${\cal B}_*$, because of the arrival of more packets than those that can be delivered. We investigate those values of $\mu$ near the
onset of congestion to reveal the effect of queues in the scaling
properties of the system. When congestion occurs, the queues
corresponding to those nodes with ${\cal B}_*$ will have always more packets that those than can be
delivered in a period $P$. That means that the number of packets
delivered by these nodes will be controlled exclusively by the
service rate $\mu$, i.e. the variance scaling with respect to the
mean flow at these nodes will be again fitted by $\alpha=1/2$
corresponding to the exponential service distribution. Close to the
onset of congestion we approach the situation where the scaling
exponent $\alpha=1/2$ should be recovered, however the possibility
that in some periods of time the queues will be unoccupied increases
as we go away from the congested regime, thus a new transition in the scaling
exponent as a function $\mu$ is expected. In Fig. \ref{mu} we plot
the scaling exponent transition as a function of $\mu$ for a fixed
value of  $\lambda^{ef}_{*}=1/3$. In this situation the onset of
congestion is determined by the critical value $\mu_c = 1/3$.  Note
that for values below $\mu_c$ some nodes of the network collapse and
then gradually the rest of the nodes in the network. In this region,
shadowed area of Fig. \ref{mu} the system enters the congestion regime progressively. The transition on the scaling exponent depicted in
Fig. \ref{mu} is also affected by the time window size $P$, we plotted
the transition for $T=10^2$, $10^3$ and $10^4$. We observe that as
$P$ increases, the transition becomes sharper. Indeed in the limit of $T\rightarrow \infty$  we conjecture that the transition could be discontinuous, and could reflect a first order phase transition \cite{stanley} as observed in other traffic models \cite{yamir2}, although we can not claim that this discontinuity will occur sharply from $1$ to $1/2$.

\begin{figure}[t]
 \epsfig{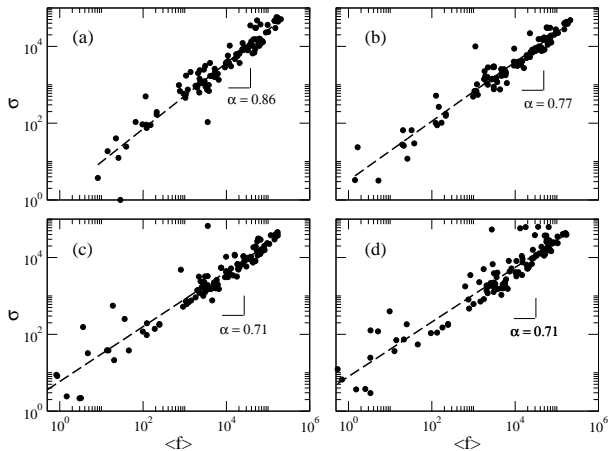}
  \caption{Scaling relations between $\sigma$ and $\langle f \rangle$ for the 112 Abilene backbone router Interfaces. Analysis performed during (a) two days, (b) one week, (c) one month and (d) two months, finishing all them in November 15th of 2005. The time window size $P$ is fixed to 5 minutes.}
\label{int1}
\end{figure}

Up to now, we have show that a simple traffic model where the injection of packets to the system follows a Poisson distribution, can account for different scaling exponents $\alpha$ depending on the parameters $\lambda$, $\mu$, $S$ and the time period $P$ were the statistics are performed. These results lead us to suspect that the scaling of fluctuations in real systems must be affected by these parameters as well. This cast doubts on the universality predicted in \cite{menezes04a}. Indeed, this non-universality has been also claimed in the exponent of fluctuations when studying the data flow between stocks in NYSE market \cite{eisler}. To corroborate our doubts about universality on the scaling of fluctuations in complex networks, we have studied the Internet traffic between routers of the Abilene backbone network \cite{footnote3} that are part of the data also used in \cite{menezes04a}. We collected data from the 112 available router interfaces (links). We gather information of the number of packets that exit through each router interface between September 15th and November 15th of 2005, at intervals of 5 minutes. The scaling $\sigma\sim \langle f \rangle^{\alpha}$ shows exponents that range from $\alpha=0.71$ to $\alpha=0.86$, significantly different from the exponent $1/2$ presented in \cite{menezes04a}. The interpretation of these exponents in the context of our stylized model is that the Abilene backbone is far from the onset of congestion for the interface with maximum algorithmic betweenness, and seems compatible with the mean rate of utilization of the interfaces in this backbone that is usually below $30\%$.

Summarizing, we have presented a simple model of traffic in complex networks that capture the essential parameters governing the dynamical process. The model shows a scaling relationship between $\sigma$ and $\langle f \rangle$ whose exponent depends on the parameters considered as well as on the time window in which the statistics are performed. Moreover we have shown that the corresponding exponent for the scaling of fluctuations in the Internet Abilene backbone network is different from $1/2$ as stated in previous works, corroborating by exclusion that the universality on the scaling of fluctuations in complex networks should be questioned. 

\begin{acknowledgments}
  We thank R. Guimer\`{a} for inspiration, discussion and helpful comments. We also thank M. Boguna, L. Danon, A. Diaz-Guilera, Y. Moreno and M.A. Serrano for helpful discussion and comments. This work has been supported by DGES of the Spanish Government Grant No. BFM-2003-08258.
\end{acknowledgments}


\begin{thebibliography}{21}
\bibitem{footnote1} Recent studies \cite{broido} show that backbone traffic appears to be well described by Poisson packet arrival.
\bibitem{footnote2} The communication process in the case of a M/M/1
queuing system for each node in a complex network is well described
by the so-called Jackson networks \cite{jackson}
\end{thebibliography}
\end{document}